\documentstyle[prd,aps,floats,psfig]{revtex}

\def\met{\mbox{${\hbox{$E$\kern-0.6em\lower-.1ex\hbox{/}}}_T$}} 
\def\D0{D\O}                            
\def\d0draft{}
\def\err#1#2#3 {{\it Erratum} {\bf#1},{\ #2} (19#3)}
\def\ib#1#2#3 {{\it ibid.} {\bf#1},{\ #2} (19#3)}
\def\nc#1#2#3 {Nuovo Cim. {\bf#1} ,#2(19#3)}
\def\nim#1#2#3 {Nucl. Instr. Meth. {\bf#1},{\ #2} (19#3)}
\def\np#1#2#3 {Nucl. Phys. {\bf#1},{\ #2} (19#3)}
\def\pl#1#2#3 {Phys. Lett. {\bf#1},{\ #2} (19#3)}
\def\prev#1#2#3 {Phys. Rev. {\bf#1},{\ #2} (19#3)}
\def\prl#1#2#3 {Phys. Rev. Lett. {\bf#1},{\ #2} (19#3)}
\def\rmp#1#2#3 {Rev. Mod. Phys. {\bf#1},{\ #2} (19#3)}
\def\zp#1#2#3 {Zeit. Phys. {\bf#1},{\ #2} (19#3)}
\def\pp{$p\bar{p}$}

\def\sb{$\widetilde b$}
\def\msb{$m_{\tilde b}$}
\begin{document}


%
\preprint{FNAL-PUB-99-046-E}
\title{Search for  bottom squarks in \pp ~collisions at $\sqrt{s} =$1.8 TeV}
\date{submitted  to PRD March 12, 1999}

\author{                                                                      
B.~Abbott,$^{43}$                                                             
M.~Abolins,$^{40}$                                                            
V.~Abramov,$^{18}$                                                            
B.S.~Acharya,$^{11}$                                                          
I.~Adam,$^{42}$                                                               
D.L.~Adams,$^{52}$                                                            
M.~Adams,$^{27}$                                                              
S.~Ahn,$^{26}$                                                                
V.~Akimov,$^{16}$                                                             
G.A.~Alves,$^{2}$
% ...
\\                                                                            
\vskip 0.30cm                                                                 
\centerline{(D\O\ Collaboration)}                                             
\vskip 0.30cm                                                                 
}                                                                             
\address{                                                                     
\centerline{$^{1}$Universidad de Buenos Aires, Buenos Aires, Argentina}       
\centerline{$^{2}$LAFEX, Centro Brasileiro de Pesquisas F{\'\i}sicas,         
                  Rio de Janeiro, Brazil}                                     
\centerline{$^{3}$Universidade do Estado do Rio de Janeiro,                   
                  Rio de Janeiro, Brazil}                                     
\centerline{$^{4}$Institute of High Energy Physics, Beijing,                  
                  People's Republic of China}                                 
\centerline{$^{5}$Universidad de los Andes, Bogot\'{a}, Colombia}             
\centerline{$^{6}$Universidad San Francisco de Quito, Quito, Ecuador}         
\centerline{$^{7}$Institut des Sciences Nucl\'eaires, IN2P3-CNRS,             
                  Universite de Grenoble 1, Grenoble, France}                 
\centerline{$^{8}$DAPNIA/Service de Physique des Particules, CEA, Saclay,     
                  France}                                                     
\centerline{$^{9}$Panjab University, Chandigarh, India}                       
\centerline{$^{10}$Delhi University, Delhi, India}                            
\centerline{$^{11}$Tata Institute of Fundamental Research, Mumbai, India}     
\centerline{$^{12}$Kyungsung University, Pusan, Korea}                        
\centerline{$^{13}$Seoul National University, Seoul, Korea}                   
\centerline{$^{14}$CINVESTAV, Mexico City, Mexico}                            
\centerline{$^{15}$Institute of Nuclear Physics, Krak\'ow, Poland}            
\centerline{$^{16}$Institute for Theoretical and Experimental Physics,        
                   Moscow, Russia}                                            
\centerline{$^{17}$Moscow State University, Moscow, Russia}                   
\centerline{$^{18}$Institute for High Energy Physics, Protvino, Russia}       
\centerline{$^{19}$University of Arizona, Tucson, Arizona 85721}              
\centerline{$^{20}$Lawrence Berkeley National Laboratory and University of    
                   California, Berkeley, California 94720}                    
\centerline{$^{21}$University of California, Davis, California 95616}         
\centerline{$^{22}$University of California, Irvine, California 92697}        
\centerline{$^{23}$University of California, Riverside, California 92521}     
\centerline{$^{24}$Florida State University, Tallahassee, Florida 32306}      
\centerline{$^{25}$University of Hawaii, Honolulu, Hawaii 96822}              
\centerline{$^{26}$Fermi National Accelerator Laboratory, Batavia,            
                   Illinois 60510}                                            
\centerline{$^{27}$University of Illinois at Chicago, Chicago,                
                   Illinois 60607}                                            
\centerline{$^{28}$Northern Illinois University, DeKalb, Illinois 60115}      
\centerline{$^{29}$Northwestern University, Evanston, Illinois 60208}         
\centerline{$^{30}$Indiana University, Bloomington, Indiana 47405}            
\centerline{$^{31}$University of Notre Dame, Notre Dame, Indiana 46556}       
\centerline{$^{32}$Purdue University, West Lafayette, Indiana 47907}          
\centerline{$^{33}$Iowa State University, Ames, Iowa 50011}                   
\centerline{$^{34}$University of Kansas, Lawrence, Kansas 66045}              
\centerline{$^{35}$Louisiana Tech University, Ruston, Louisiana 71272}        
\centerline{$^{36}$University of Maryland, College Park, Maryland 20742}      
\centerline{$^{37}$Boston University, Boston, Massachusetts 02215}            
\centerline{$^{38}$Northeastern University, Boston, Massachusetts 02115}      
\centerline{$^{39}$University of Michigan, Ann Arbor, Michigan 48109}         
\centerline{$^{40}$Michigan State University, East Lansing, Michigan 48824}   
\centerline{$^{41}$University of Nebraska, Lincoln, Nebraska 68588}           
\centerline{$^{42}$Columbia University, New York, New York 10027}             
\centerline{$^{43}$New York University, New York, New York 10003}             
\centerline{$^{44}$University of Rochester, Rochester, New York 14627}        
\centerline{$^{45}$State University of New York, Stony Brook,                 
                   New York 11794}                                            
\centerline{$^{46}$Brookhaven National Laboratory, Upton, New York 11973}     
\centerline{$^{47}$Langston University, Langston, Oklahoma 73050}             
\centerline{$^{48}$University of Oklahoma, Norman, Oklahoma 73019}            
\centerline{$^{49}$Brown University, Providence, Rhode Island 02912}          
\centerline{$^{50}$University of Texas, Arlington, Texas 76019}               
\centerline{$^{51}$Texas A\&M University, College Station, Texas 77843}       
\centerline{$^{52}$Rice University, Houston, Texas 77005}                     
}                                                                             
\maketitle
\begin{abstract}

We report on  a search for bottom squarks
(\sb) produced in $p\bar{p}$ collisions at $\sqrt s = 1.8$ TeV
using the \D0 detector at Fermilab. Bottom
squarks are assumed to be produced in pairs and 
to decay to the lightest supersymmetric
particle (LSP) and a $b$ quark with a branching fraction of
100\%. The LSP is assumed to be the
lightest neutralino and stable. We set limits 
on the production
cross section as a function of \sb  ~mass 
and LSP mass. 
\end{abstract}

\twocolumn
%

Supersymmetry (SUSY) is a hypothetical
fundamental space-time symmetry relating
bosons and fermions \cite{theory}. Supersymmetric extensions to
the standard model (SM) feature as yet undiscovered supersymmetric partners for
every SM particle. The scalar quarks (squarks) $\widetilde q_L$ 
and $\widetilde q_R$ are the partners of the
left-handed and right-handed quarks, respectively. These
are weak eigenstates, and can mix to form the
mass eigenstates, with $\widetilde q_1 = \widetilde q_L$cos$\theta +
\widetilde q_R$sin$\theta$ for the lighter squark, and the orthogonal
combination for the heavier squark $\widetilde q_2$.  In most
SUSY models, the masses of the squarks
are approximately degenerate. But in some models,
the lighter top and bottom squarks  could have
a lower mass than the other squarks because of the high mass values
of the top and bottom quarks. In particular, lighter
bottom squarks could arise for large values of tan$\beta$, the ratio
of the vacuum expectation values of the two Higgs fields
in the minimal supersymmetric standard model. 

We report the results of a mixing-independent search for 
bottom squarks produced in
\pp ~collisions at $\sqrt s = 1.8$ TeV. 
Squarks are produced in pairs by QCD processes with
the production cross section depending on the mass of
the squark but
not on the mixing angle $\theta$.
We search for events where both squarks decay to
the lightest neutralino $\widetilde\chi_1^0$ via
$\widetilde b \rightarrow \widetilde\chi_1^0 + b$ and assume that the
$\widetilde\chi_1^0$ is the lightest supersymmetric particle (LSP)
and stable. 
This should be the dominant decay channel
provided that the mass of the squark (\msb) is larger than
the combined masses of the $b$ quark and LSP ($m_{\text{LSP}}$);
therefore we assume its branching fraction is 100\%. 
This yields a final state consisting of
two $b$ quarks and two unobserved stable particles
resulting in missing transverse energy (\met)
in the detector.
In this paper, we give
limits on the squark pair production cross section
for different values of \msb~ and $m_{\text{LSP}}$.
Limits on the cross section are used to exclude
a region in the ($m_{\text{LSP}}$, \msb) plane.
Limits \cite{lep}  from the CERN $e^+e^-$ collider (LEP)
experiments depend on
the $Z/\gamma$-to-squark coupling, which
is a function of the mixing angle.  For maximal
coupling, the LEP exclusion region can extend
to the kinematic maximum; for example, to about 85 GeV/$c^2$
at $\sqrt{s}=183$ GeV.

The data used for our analysis were collected during
1992--1996 by the \D0
detector\cite{dzero} at the Fermilab Tevatron Collider.
The \D0 detector
is composed of three major systems: an inner detector for
tracking charged particles, a uranium/liquid argon calorimeter for
measuring electromagnetic and hadronic energies, and a muon spectrometer
consisting of a magnetized iron toroid and three layers of drift tubes.
The detector measures jets with an energy resolution of approximately   
$\sigma /E = 0.8/\sqrt E$ ($E$ in GeV) and muons with a momentum
resolution of 
$\sigma /p=[(\frac{0.18(p-2)}{p})^2 + (0.003 p)^2]^{1/2}$  ($p$ in GeV$/c$).
\met~is determined by
summing the calorimeter and muon transverse energies, and is measured with
a resolution of $\sigma$ = 1.08 GeV + 0.019$\cdot (\Sigma |E_T|)$ \cite{topprd}.

Four channels are combined to set limits on the production of 
bottom squarks. The first required a \met~and jets topology.
This channel was previously used to set limits on the mass of
the top squark, which was assumed to decay 
$\widetilde t \rightarrow \widetilde\chi^0_1 +c$ \cite{stop}.
The other three channels in addition required that at least one jet has an
associated muon, thereby tagging $b$ quark decay, and were
used to set limits on a charge 1/3 third
generation leptoquark for the decay $LQ\rightarrow \nu_\tau + b$ \cite{lq3}. 
We use identical data samples and event selections 
for the bottom squark limits 
presented in this paper.  For all channels, the
presence of significant \met~is used to identify the non-interacting
LSPs. Figure 1 shows the expected \met~distribution for
two values of \msb~and different $m_{\text{LSP}}$ \cite{monte}. Our requirement that
\met$>35-40$ GeV reduces the acceptance for
small values of the mass difference \msb --$m_{\text{LSP}}$.
Backgrounds arise from events where 
neutrinos produce significant \met; for example, in $W$+jets
events, where $W\rightarrow l\nu$.

\begin{figure}\vbox{
\centerline{
\psfig{figure=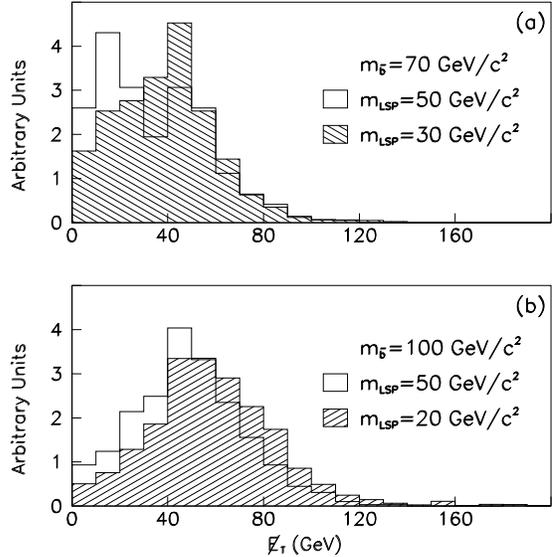,width=3.3in}}
\caption{ The expected distributions of \met~for  
\msb ~ values of
 70 (a) and  100 (b)  GeV/$c^2$, for the indicated values of $m_{\text{LSP}}$ 
[7].}
}
\end{figure}

Events for the \met+jets channel were collected using a trigger that
required \met$>35$ GeV.
The offline analysis required  two jets ($E_T^{\rm jet}>30$
GeV), \met$>40$ GeV, and no isolated electrons or muons.
Events had to have only one primary vertex
to assure an unambiguous calculation of \met.
To eliminate QCD
backgrounds, additional
cuts were made on the angles between the two jets, and between
jets and the direction of the \met. Data with 
an integrated luminosity
of 7.4 pb$^{-1}$, satisfying the above selection criteria,
yielded three candidate events.
Background was  estimated to be $3.5\pm 1.2$ events, with
$3.0\pm 0.9$ events from  $W$ boson decays and $0.5\pm 0.3$
events from $Z$ boson decays \cite{stop}.
 
The trigger for the muon channels required either two
low-$p_T$ muons ($p_T^\mu >3.0$ GeV/$c$), or a single
low-$p_T$ muon and a jet with $E_T>10$ GeV, or a
 high-$p_T$ muon ($p_T^\mu >15$ GeV/$c$) and a jet
with $E_T>15$ GeV.
Integrated luminosities of 60.1 pb$^{-1}$, 19.5 pb$^{-1}$,
and 92.4 pb$^{-1}$ respectively were collected
using the three muon triggers.   
The offline analysis used muons   
in the pseudorapidity range $|\eta_\mu|$ $<$  1.0
and $p_T^\mu>3.5$ GeV/$c$, while jets were
required to have $E_T>10$ GeV. 
For events with two muons, each muon had to be
associated with its own jet. In single muon events,
the muon was required to be associated with a jet, and
an additional jet with $E_T>25$ GeV 
was also required. To remove  QCD  backgrounds,
events were selected with \met~$>35$ GeV and an azimuthal
angular separation between the \met~and
the nearest jet of $>0.7$ radians. 
For the single muon channels,
backgrounds from $W$ boson decays were reduced by cuts
on muon-jet correlations, while background from
top quark production was minimized
by cuts on the scalar sum of jet $E_T$. 
After imposition of all selection
criteria, two events remained in the data.

We considered background contributions to the
muon channels from $t\bar{t}$ and
$W$ and $Z$ boson decays \cite{lq3}. Top quark events
have multiple $b$ quarks and \met, and we estimated that
$1.4\pm 0.5$ $t\bar{t}$ events remained in our sample.
$W$ and $Z$ events have \met~from
$W\rightarrow l\nu$ or $Z\rightarrow \nu\bar{\nu}$. They can
also have 
muons near jets that can mimic $b$ quark decays when 
a prompt muon overlaps a jet, or a jet fragments into
a muon via a $c$ quark or a $\pi/K$ decay. We estimated there
were $1.0\pm 0.4$ $W$ boson events and $0.1\pm 0.1$ $Z$ boson
events in the sample.  
The total background for the muon channels
was therefore $2.5\pm 0.6$ events.
 
Combining the four channels yields five events, 
with a total estimated background of $6.0\pm 1.3$ events.
We set limits on the cross
section by combining the detection efficiencies and integrated
luminosities for the different channels.
We calculate the detection efficiency using Monte Carlo (MC) 
generated acceptances \cite{monte}, multiplied by trigger and
reconstruction efficiencies obtained from data [5,6]. 
The total efficiencies for different squark and neutralino masses
are summarized in Table I. Using  a 
muon to tag $b$ quark decays reduces the
efficiency for those channels, but their higher integrated
luminosities yield  a sensitivity comparable to that of
the \met+jets channel.
Including systematic errors
and statistics for the MC, the total uncertainty on the combined efficiency varies 
between 8.6\% and 29\%, depending on the assumed masses.
The jet energy scale dominates the systematic error
for \msb ~= 70 GeV/$c^2$, while uncertainties
on the muon trigger and reconstruction efficiency dominate
at higher squark masses. 
The 95\% confidence level (C.L.) upper limits on  
the pair production cross section are determined
using Bayesian methods, and include the
systematic uncertainty on the efficiency  and a 5.3\% uncertainty in
the integrated luminosity. The resulting upper limits  
are given in Table I for different values of
\msb~ and $m_{\text{LSP}}$. 

\begin{table}
\caption{Total efficiencies for different \msb ~and $m_{\text{LSP}}$ values for
the four channels, and     
95\% C.L. limits on the production cross section obtained by
combining all channels.}
\begin{tabular}{c c | c c c c | c}
\msb& $m_{\text{LSP}}$   & \multicolumn{4}{c|}{Total efficiency ($\times 10^{-3}$)} &  $\sigma $ limit  \\
 \multicolumn{2}{c|}{(GeV/$c^2$)}  & \met +  & dimuon  & \multicolumn{2}{c|}{ single muon} &  (pb) \\ 
 &  & jets &    &  low-$p_T$  &  high-$p_T$  &  \\ 
\hline
~70 & 30 &18 &0.13& ~2.2 & 0.3 & ~32~~ \\
~70 & 50 &~4 &0.02 & ~0.6 & 0.1 & 245~~ \\
~85 & 40 & 29&0.20 & ~3.9 & 0.6 & ~18.8 \\
~85 & 60 & 11&0.04 & ~1.0 & 0.1& ~84~~  \\
100 & 20 & 43&0.50 & ~9.5 & 1.9 & ~~9.3 \\
100 & 40 & 34&0.27 & ~7.0 & 1.3 & ~12.6 \\
100 & 50 & 30 &0.30 & ~5.8 & 1.0 & 14.7 \\
115 & 40 & 51 &0.54 & 10.9 & 2.0 & ~~8.0 \\
\end{tabular}
\label{tab:table2}
\end{table}

We use the program {\sc prospino} \cite{NLO} to
calculate the  bottom squark pair
production cross section as a function of \msb. The cross section
is evaluated assuming a renormalization
scale $\mu = m_{\widetilde b}$. The program includes next-to-leading 
order diagrams, and uses {\sc cteq4m} parton distribution functions \cite{cteq}.
For any given \msb, we determine the value of $m_{\text{LSP}}$
where our 95\% C.L. limit intersects the theoretical
cross section. The excluded region in the
($m_{\text{LSP}}$,\msb) plane is shown in Fig. 2.
We exclude values of \msb ~ below 115 GeV/$c^2$ for
$m_{\text{LSP}}<20$ GeV/$c^2$. For \msb ~= 85 GeV/$c^2$,
we exclude the region with $m_{\text{LSP}}< 47$ GeV/$c^2$. 
Also shown are limits \cite{lep}
from ALEPH  for $\sqrt{s} = 181-184$ GeV.
 For most allowable values of $m_{\text{LSP}}$, they exclude the
region with \msb ~$<83$ GeV/$c^2$, assuming maximal coupling ($\theta=0^o$)
\cite{aleph}.

\begin{figure}\vbox{
\centerline{
\psfig{figure=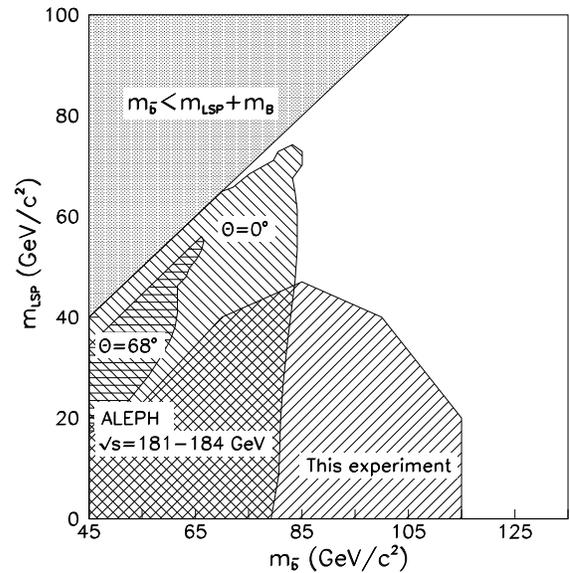,width=3.3in}}
\caption{The  95\% C.L. exclusion contour
in the ($m_{\text{LSP}}$,\msb)  plane. Also shown
are the results from the ALEPH experiment at LEP
for minimal ($\theta=68^o$) and maximal ($\theta=0^o$) coupling [2]. }
}
\end{figure}

In conclusion, we observe five candidate events 
consistent with the final state $b\bar{b}+$\met.
We estimate that $6.0\pm 1.3$
events are expected  from $t\bar{t}$ and $W$ and $Z$ boson production,
and find no excess of events that can be attributed
to bottom squark
production.
We interpret our result as an excluded
region in the ($m_{\text{LSP}}$,\msb) plane. 
This result is independent of the
mixing  between $\widetilde b_L$ and $\widetilde b_R$.  

We thank S.P. Martin and M. Spira for their assistance.
We thank the Fermilab and collaborating institution staffs for
contributions to this work and acknowledge support from the 
Department of Energy and National Science Foundation (USA),  
Commissariat  \` a L'Energie Atomique (France), 
Ministry for Science and Technology and Ministry for Atomic 
   Energy (Russia),
CAPES and CNPq (Brazil),
Departments of Atomic Energy and Science and Education (India),
Colciencias (Colombia),
CONACyT (Mexico),
Ministry of Education and KOSEF (Korea),
and CONICET and UBACyT (Argentina).

\end{document}